\documentclass[10pt]{article}
\usepackage{amsfonts}
\usepackage{epsfig}
\begin{document}

\title{Modelling Complex Networks: Cameo Graphs And Transport Processes}
\author{ Philippe Blanchard and Dimitry Volchenkov
\vspace{0.5cm}\\
{\small \it Bielefeld-Bonn Stochastic Research Center (BiBoS)},\\
{\small\it Bielefeld University, Postfach 100131, 33501, Bielefeld, Germany}\\
{\small\it Email: volchenk@physik.uni-bielefeld.de}}

\date{\today}
\maketitle
\begin{abstract}
We discuss a model accounting for the creation and development of transport networks based on the Cameo principle which refers to the idea of distribution of resources, including land, water, minerals, fuel and wealth. We also give an outlook of the use of random walks as an effective tool for the investigation of network structures and its functional segmentation. In particular, we have studied the complex transport network of Venetian canals by means of random walks. 
\end{abstract}

\vspace{0.5cm}

\begin{flushleft}
\textbf{Presentation for the volume:}
\textsl{ The challenge of complex network modelling calls for the more realistic heuristic principles that could catch the main features of network creation and development. In the Cameo model which refers to the idea of distribution of resources, including land, water, minerals, fuel and wealth, the local attractiveness of a site determining the creation of new spaces of motion in that is specified by a real positive parameter $\omega >0$.  We have described a possible mechanism for the emergence and development of complex transport networks based on the Cameo principle.
Sustained movement patterns are generated by a subset of automorphisms of the graph spanning the transport network of can be naturally interpreted as random walks. Random walks assign absolute scores to all nodes of a graph and embed space syntax into Euclidean space. Namely, every route of a transport network can be represented by a vector in Euclidean space which length quantifies the segregation of the route from the rest of the graph. We have empirically observed that the distribution of lengths over the edge connectivity in the spatial network of Venetian canals exhibits scaling invariance phenomenon. The method is applicable to any transport network.}
\end{flushleft}

\vspace{0.5cm}


\section{Introduction}
\label{sec:Introduction}
 \noindent

Physics (Greek: $\mathrm{\varphi \upsilon\sigma\iota\varsigma}$
[phusis], nature) is the branch of science concerned with the
characterization of universal laws of Nature portraying its logically
ordered picture in agreement with experience. Theoretical physics is
closely related to mathematics, which provides a language for physical
theories and allows for a rationalization of thought by making it possible to
formulate these
laws in terms of mathematical relations.
Physicists study a wide variety of phenomena creating new interdisciplinary
research fields by applying theories and methods originally developed in
physics in order to solve problems in economics, social science, biology,
medicine, technology, etc. In their turn, these different branches of
science inspire the invention of new concepts in physics. A basic tool
of analysis, in such a context, is the mathematical theory of complexity
 concerned with the study of complex systems including human economies,
 climate, nervous systems, cells and living things, including human beings,
  as well as modern energy or communication infrastructures which are all
  networks of some kind.

Complex systems appear as a result of the interplay between
Topology determined by a connected graph, Dynamics described by
the operators invariant with respect to graph symmetry, and
properties of embedding (Euclidean) space specified by a set of
measures and weights assigned to elements of the graph. In the
context of complex networks theory created by physicists, the
non-trivial topological structure of large networks is
investigated by means of various statistical distributions. The
structure and the properties of complex networks essentially
depend on the way how nodes get connected to each other. Random
graphs with a scale free distribution for the degree seem to
appear very frequently in a great variety of real life situations
like the World- Wide Web, the Internet, social networks,
linguistic networks, citation networks and biochemical networks.

In most of complex networks emerging in society and technology, each
node has a feature which attracts the others. In a class of simple
 models proposed in \cite{Cameo}, the network dynamics can be described
  in terms of property of the node and the affinity other nodes have
   towards that property (Cameo graphs). Networks built accordingly
    to this principle have a degree distribution with a power law
     tail, whose exponent is determined only by the nodes with the
     largest affinity values. It appears that the extremists lead
     the formation process of the network and manage to shape the
     final topology of the system.

The exceptional events play a crucial role in the formation of network
 structures \cite{BlanchardKruger:1992}.
The dynamics of some vertices — the "hubs" which have an extremely
high number of connections to other vertices is of primary
importance for complex networks. These networks are generally
"scale-free"; in other words, they exhibit architectural and
statistical stability as the degree distribution grows. A class of
probabilistic model for a system at a threshold of instability has
been studied in \cite{Volchenkov:2003}. The distribution of
residence times below the threshold characterizes the properties
of such a system. Being at a threshold of instability, the system
can induce various types of random graphs and the scale free
random graphs among others \cite{Volchenkov:2002}. The
priority-based scheduling rules in single-stage queuing systems
(QS) also generate fat tail behavior for the task waiting time
distributions induced by the waiting times of very low priority
tasks that stay unserved almost forever as the task priority
indices are "frozen" in time \cite{Hongler:2007}. The task waiting
time distributions have been studied for a population-type model
with an age structure and a QS with deadlines assigned to the
incoming tasks, which is operated under the
"earliest-deadline-first" policy. As the aging mechanism
ultimately assigns high priority to any long waiting tasks, fat
tails cannot find their origin in the scheduling rule alone.

Graphs obtained by successive creation and elimination of edges
into small neighborhoods of the vertices evolve towards small
world graphs with logarithmic diameter, high clustering
coefficients and a fat tail distribution for the degree \cite{Ruschhaupt}.
It is important to note that it was only local edge formation
 processes that rise small worlds, no preferential attachment
  was used. Simple edge generation rules based on an inverse
  like mass action principle for random graphs over a
   structured vertex set, under very weak assumptions
    on the structure generating distribution, also yield
    a scale free distribution for the degree \cite{Sigurue}.
    A local search principle important in many social
    applications, "my friends are your friends" have
    also been introduced and studied; networks generated
    in accordance to such a principle have essentially
     high clustering coefficients.

Although investigations into the statistical properties of
 graphs such as a heavy-tail in the degree distribution of
 nodes could uncover their hierarchical structure, they are
 futile if the detailed information on the structure of graphs
  is of primary interest since many graphs characterized by
  similar statistics of node degrees and shortest path lengths
  can be of dramatically different structures. The structure
  and symmetry of graphs play the crucial role in behavior of
   dynamical systems defined on that. It was clearly
   demonstrated in epidemiological research describing the
   dynamics of sexually transmitted diseases, the Human
   Immune Deficiency Virus (HIV) and AIDS, in particular
    \cite{BlanchardKruger:1992}.
 Mathematical modelling on the spread of sexually transmitted
 diseases \cite{Bolz:1990}-\cite{BlanchardBolz:1990}
 studied on various random graphs displays the importance of
  critical parameters such as the transmission probability
   and edge creation probability for the epidemic spreading.

 It has been found that the epidemic spreading in scale-free
 networks is very sensitive to the statistics of degree distribution,
  the effective spreading rate, the social strategy used by individuals
   to choose a partner, and the policy of administrating a cure to an
    infected node \cite{Volchenkova}. Depending on the interplay of
     these four factors, the stationary fractions of infected population
      as well as the epidemic threshold properties can be essentially different.
      For a model of scale-free graphs with biased partner choice that
      knowing the exponent for the degree distribution is in general not
      sufficient to decide epidemic threshold properties for exponents
      less than three \cite{UNESCO}. Absence of epidemic threshold happens
      precisely when a positive fraction of the nodes form a cluster of
       bounded diameter. Probably, it is impossible to obtain a simple
        immunization program that can be simultaneously effective for
         all types of scale-free networks \cite{Volchenkova}.
A similar approach can be applied in order to study social diseases
 like corruption. It has been investigated in \cite{Krueges:2005} as a
 generalized epidemic process on the graph of social relationships.
 Corruption is characterized by a strong nonlinear dependence of the
 transmission probability from the local density of corruption and
  the mean field influence of the overall corruption in the society.
   Network clustering and the degree-degree correlation play an essential
    role in these types of dynamics. In particular, it follows that
    strongly hierarchically organized societies are more vulnerable
    to corruption than democracies. A similar type of modelling can
     be applied to other social contagion spreading processes like
      opinion formation, doping usage, social disorders or innovation
      dynamics. An agent-based model of factual communication in social
       systems, drawing on concepts from Luhmann's theory of social
       systems \cite{Luhman} has been studied in \cite{Barber_et_al:2006}.
        The agent communications are defined by the exchange of distinct
         messages. Message selection is based on the history of the
         communication and developed within the confines of the problem
          of double contingency. We have examined the notion of
           learning in the light of the message-exchange description.

Topology plays the primary role in the dynamical processes which have place on networks.
The investigations in transitions to spatio-temporal intermittency
in random network of coupled Chat\'{e} –- Manneville maps
\cite{Sequeira} show that spatiotemporal intermittency occurs for
some intervals or windows of the values of the network connectivity,
 coupling strength, and the local parameter of the map. Within the
  intermittency windows, the system exhibits periodic and other
   nontrivial collective behaviors. Genetic regulatory networks
    constitute an important example of dynamical systems defined
     on graphs. Local dynamics of network nodes exhibits multiple
      stationary states and oscillations depending crucially upon
      the global topology of a 'maximal' graph (comprising of all
       possible interactions between genes in the network)
       \cite{Genes}. The long time behavior observed in the
       network defined on the homogeneous 'maximal' graphs is
       featured by the fraction of positive interactions
        (activations) allowed between genes. In networks
        defined on the inhomogeneous directed graphs depleted
        in cycles, no oscillations arise in the system even if
         the negative interactions (inhibitions) in between
          genes present therein in abundance. Local dynamics
           observed in the inhomogeneous scalable regulatory
            networks is less sensitive to the choice of initial conditions.

In mathematics, the automorphism groups of a graph are studied.
 They characterize its symmetries, and are therefore very useful
  in determining certain of its properties. In particular, the
   Euclidean metric related to dynamics can be defined on some
   graphs by means of linear operators remaining invariant under
   the permutations of nodes and satisfying some conservation
    properties. These operators describe certain dynamical
     processes defined on
graphs such as random walks and diffusions. We have studied
transport through generalized trees in \cite{VB_Stat:2007}.
Trees contain the simple nodes and super-nodes, either
 well-structured regular subgraphs or those with many triangles.
 We observe super-diffusion for the highly connected nodes while
  it is Brownian for the rest of the nodes. Transport within a
  super-node is affected by the finite size effects vanishing as
   $N\to\infty$. For a space of even dimensions, $d = 2, 4, 6\ldots$,
   the finite size effects break down the perturbation theory at small
    scales and can be regularized by using a heat-kernel expansion.
Diffusion processes and Laplace operators related to them can be
used in order to investigate the structure of networks in the spirit
 of spectral graph theory. In \cite{cities:2007}, different models
  of random walks on the dual graphs of compact urban structures are
  considered. Dual graphs have been widely used in the framework of
   space syntax theories \cite{Hillier:1999} for the analysis of
    spatial configurations. The general idea is that spaces can be
     broken down into components, analyzed as networks of choices,
      and then represented as maps and graphs that describe the
       relative connectivity and integration of those spaces.
        From these components it is thought to be possible to
         quantify and describe how easily navigable any space
          is, useful for the design of museums, airports,
          hospitals, and other settings where way finding is
           a significant issue. Space syntax has also been
           applied to predict the correlation between spatial
           layouts and social effects such as crime, traffic
            flow, sales per unit area, etc. Analysis of
            access times between streets performed in
             \cite{cities:2007} helps to detect the city modularity.

The aim of the present paper is twofold. First, we discuss a model
which accounts for the creation and development of  transport
networks basing on the Cameo principle \cite{Cameo} which refers to
the idea of distribution of resources, including land, water,
minerals, fuel and wealth in general (see Sec.~\ref{sec:Cameo}).
Second, we give an outlook of the use of random walks as an
effective tool for the investigation of network structures and its
functional segmentation (Sec.~\ref{sec:Arguments}). In the
Sec.~\ref{sec:Examples}, we consider two examples of graphs (the
modelling example of the Petersen graph and the spatial network of
Venetian canals) and analyze their properties. We conclude in the
last section.

\section{The Cameo principle and the origin of transport networks}
\label{sec:Cameo}
\noindent

Among the classical models in which the degree distribution
of the arising graph
satisfies a power-law is
the graph generating algorithms
based on  the  preferential attachment approach firstly proposed by H. Simon \cite{Simon:1955}. 
Within preferential
   attachment algorithms, the growth of a network starts
   with an initial graph of $n_0\geq 2$ nodes such that
   the degree of each node in the initial network is at
   least 1.
The celebrated Barab\'{a}si-Albert model \cite{BA} have been
 proposed in order to model the emergency and growth
  of scale-free complex networks.
 New nodes in the model \cite{BA} are
added to the network one at a time. Each
 new node is connected to $n$ of the existing with a
  probability that is biased being proportional
   to the number of links that the existing node already has,
\begin{equation}
\label{Bas}
p_i=\frac {\deg(i)}{\sum_{j=1}^N\deg(j)}.
\end{equation}
It is clear that the
nodes of high degrees tend to quickly accumulate even more
 links representing a strong {\it preference choice} for
 the emerging nodes, while nodes with only a few links
 are unlikely to be chosen as the destination for a new
 link.
The preferential attachment forms
 a positive feedback
loop in which an initial random degree variation is magnified with
time, \cite{BA2}. It is fascinating that the expected degree
distribution in the graph generated in accordance to the algorithm
proposed in \cite{BA} asymptotically approaches the cubic
hyperbola,
\begin{equation}
\label{cubichyperbola}
\Pr\left[i\in \mathfrak{G}|\deg(i)=k\right]\simeq \frac 1{k^3}.
\end{equation}
It is however obvious that
the mechanisms
governing the creation and development of transport networks
certainly do not
follow such a simple preferential
 attachment principle as that discussed in \cite{BA}.
Indeed, nowadays the new transportation routes are usually created
as a result of the subdivision or redevelopment of an existing
transport network. Appearing due to the complicated trade-off
processes between multiple objectives, they can hardly be planed
in such a way as to meet
 the transportation routes that already have the ever
maximal number of junctions
with other  routes in the network.

The challenge of complex network modelling calls for the more
realistic heuristic principles that could catch the main features
of network creation and development. It is clear that a prominent
model should take into account
the structure of embedding physical space:
the size and shape of landscape, and the
 local land use patterns if a city transportation network is considered.
A suitable algorithm
describing the
 development of
complex networks which
takes into account
the properties of the surrounding place
has been recently  proposed
in \cite{Cameo}.
It is called the {\it Cameo-principle} having
in mind the attractiveness,
rareness and beauty of the small medallion with a
 profiled head in relief called
Cameo. It is exactly their rareness and beauty which
gives them their high
value.

In the  Cameo model \cite{Cameo}, the local attractiveness of a
site determining the creation of new spaces of motion in that is
specified by a real positive parameter $\omega >0$. Indeed, it is
rather difficult if ever be possible to estimate exactly
 the actual value $\omega(i)$ for any
 site $i\in \mathfrak{G}$
in the urban pattern, since such an estimation can be referred to
both the {\it local believes} of city inhabitants
 and may be to the {\it cultural context}
 of the site
 that may
vary over the
different nations, historical epochs, and
even over the certain groups of population.

Therefore, in the framework of the probabilistic approach,
 it seems natural to consider the
value $\omega$ as a real positive
 independent {\it random variable}
distributed over the
vertex set of the graph representation of the
site uniformly
 in accordance to
a smooth monotone decreasing probability
density function
$f \left( \omega \right)$.
Let us suggest
that there is just a few
distinguished
sites which are much more
attractive then an
average one in the city,
so that the density function $f$
has a right tail for
large $\omega\gg \bar{\omega}$ such that
  $f(\omega)\ll f(\bar{\omega})$.

Each newly created space of motion $i$ (represented by a node in
the dual city graph $\mathfrak{G}(N)$ containing $N$ nodes) may be
arranged in such a way to connect to the already
 existed space
$j\in\mathfrak{G}(N)$
depending only on its attractiveness
 $\omega(j)$ and is of the form
\begin{equation}
\Pr\left[\,
i\sim j \,\,|  \,\, \omega(j) \, \right]\,
\simeq\,
\frac{1}{N}\,\frac{1}{\left( 
f^{\alpha}\left(\omega(j)\right)+
f^{\alpha}\left(\omega(i)\right)
\right)}
\label{P1.C1.01}
\end{equation}
\noindent
with some $ \alpha \in \left( 0,1\right)$.  The assumption
(\ref{P1.C1.01}) implies that the probability to create
the new space adjacent to a space $j$ scales with
the rarity of sites characterized with the same
attractivity $\omega$ as $j$.

The striking observation under the above assumptions is the
emergence of a scale-free degree distribution independent of the
choice of distribution $f(\omega)$. Furthermore, the exponent in
the asymptotic degree distribution
becomes independent of the distribution $f(\omega)$ provided its tail,
  $f(\omega)\ll f(\bar{\omega}),$
 decays faster then any power law.

In the model of growing networks proposed  in \cite{Cameo}, the
initial graph $\mathfrak{G}_0$ has $N_ 0$ vertices, and a  new
vertex of attractiveness $\omega$ taken independently uniformly
 distributed in accordance to
 the given
density $f(\omega)$ is added to the already existed network at
each time click $t\in \mathbb{Z}_+$. Being associated to the
graph, the vertex establishes $k_0>0$ connections with other
vertices already present in that. All edges are formed accordingly
to the Cameo principle (\ref{P1.C1.01}).

The main result of \cite{Cameo}
is on the
probability distribution
that a randomly chosen vertex $i$
which
had joined the Cameo graph $\mathfrak{G}$
at time $\tau>0$
 with  attractiveness $\omega(i)$
amasses
precisely $k$ links from other vertices
which emerge by time $t>\tau$.
It is
important to note that
in the Cameo model the order in which the
edges are created plays a
role for the fine structure of the graphs.
The resulting degree distribution
for $t-\tau> k/k_0$
is irrelevant to
the
concrete form of $f(\omega)$ and reads as following
\begin{equation}
\label{Krueger}
\Pr\left[ \sum_{j:\,\,\, \tau(j) > \tau} 1_{i\sim j} = k\right]
\simeq
\frac{k_0^{1/\alpha}}{k^{1+1/\alpha + o(1)}}
 \ln^{1/\alpha}\left(\frac{t}{\tau}\right).
\end{equation}
In order to obtain the asymptotic probability degree distribution
for an arbitrary node as $t\to \infty$, it is
necessary to sum
(\ref{Krueger}) over all $\tau<t$
that gives
\begin{equation}
\label{Krueger2}
P(k) \simeq   \frac 1t \sum_{0<\tau< t} \frac{k_0^{1/\alpha}}
{k^{1+1/\alpha + o(1)}} \ln^{1/\alpha}\left(\frac{t}{\tau}\right)=
\frac{1}{k^{1+1/\alpha + o(1)}}.
\end{equation}
The emergence of the power law (\ref{Krueger2}) demonstrates
that graphs with a scale-free degree distribution
may appear naturally as the result of a simple edge
formation rule based on choices
where the probability to chose a vertex with
affinity parameter $\omega$ is proportional
to the frequency of appearance of that parameter.
If the affinity parameter $\omega$
is itself power law like distributed one could also
use a direct proportionality
to the value $\omega$ to get still a scale free graph.

We have described a possible mechanism for the emergence and development 
of complex transport networks based on the Cameo principle. In the forthcoming section,
we discuss the embedding of transport network into the $(N-1)-$dimensional
Euclidean space which facilitates the discovering of important nodes,
 their classification, and the coarse-graining.

\section{Mathematics of transport networks}
\label{sec:Arguments}
\noindent

Any graph representation  naturally arises as the outcome of a categorization,
when we abstract a real world system by eliminating all but one of its features
 and by  grouping together things (or places) sharing a common attribute.
For instance, the common attribute of all open spaces
 in city space syntax is that we can move through them.
All elements called
 nodes that fall into one and the same group $V$ are considered as essentially
 identical; permutations of them within the  group are of no consequence.
The symmetric group $\mathbb{S}_{N}$ consisting of all permutations of $N$
elements
($N$ being the cardinality of the set $V$) constitute therefore the symmetry group of $V$.
If we denote by $E\subseteq V\times V$ the set of ordered pairs of nodes called
edges, then  a graph is a map $G(V,E): E \to K\subseteq\mathbb{R}_{\,+}$
(we suppose that the graph has no multiple edges). If two nodes are adjacent,
 $(i,j)\in E$ we write $i\sim j$.

\subsection{The right choice for graph representation}
\label{subsec:primary-and-dual}
 \noindent

First, we establish a connection between transport flows on the graph $G$
and random walks on its dual counterpart $G^\star.$

Given a connected undirected graph $G(V,E)$, in which $V$ is the
set of nodes and $E$ is the set of edges, we introduce the traffic
function $f: E\to(0,\infty[$ through every edge $e\in E$. It then follows
  from the Perron-Frobenius theorem \cite{PerronFrobenius} that the linear equation
\begin{equation}
\label{Lim_equilibrium}
f(e)\,=\, \sum_{e'\sim\, e}\,f(e')\,\exp\left(\,-h\,\ell\left(e'\right)\,\right),
\end{equation}
where the sum is taken over all edges $e'\in E$ which have a common node with $e$,
has a unique positive solution $f(e)>0$, for every edge $e\in E$, for a
fixed positive constant $h>0$ and a chosen set of positive  {\it metric
 length} distances $\ell(e)>0$. This solution is naturally identified
 with the traffic equilibrium state of the transport network defined on
  $G$, in which the permeability of edges depends upon their lengths.
The parameter $h$  is called the volume entropy of the graph $G$, while
 the volume of $G$ is defined as the sum
\[
\mathrm{Vol}(G)\,=\,\frac 12\,\sum_{e\,\in\, E}\,\ell(e).
\]
The volume entropy $h$ is defined to be the exponential growth of the balls in
a universal covering tree of $G$ with the lifted metric, \cite{Manning}-\cite{Lim:2005}.

The degree of a node $i\in V$ is the number of its neighbors in
$G$, $\deg_G(i)=k_i$. It has been shown in \cite{Lim:2005} that
among all undirected connected graphs of normalized volume,
$\mathrm{Vol}(G)=1$, which are not cycles and for which $k_i\ne 1$ for all
nodes (no cul-de-sacs),
 the minimal  value of the volume entropy,
$\min(h)=\frac 12\sum_{i\in V}k_i\,\log\left(k_i-1\right)$  is attained
for the length distances
\begin{equation}
\label{ell_min}
\ell(e)\,=\,\frac {\log\left(\left(k_i-1\right)
\left(k_j-1\right)\right)}{2\,\min(h)},
\end{equation}
where $k_i$ and $k_j$ are the degrees of the nodes linked by $e \in E$.
It is then obvious that substituting (\ref{ell_min})
and $\min(h)$ into (\ref{Lim_equilibrium}) the
operator $\exp\left(-h \ell(e')\right)$ is given by
 a symmetric Markov transition operator,
\begin{equation}
\label{Markov_transition}
f(e)\,=\, \sum_{e'\,\sim\, e}\,\frac{f(e')}{\sqrt{\left(k_{i}-1\right)
\left(k_{j}-1\right)}},
\end{equation}
where $i$ and $j$ are the  nodes linked by $e' \in E$, and the sum in
(\ref{Markov_transition}) is taken over all
edges $e'\in E$ which share a node with $e$.
The symmetric operator (\ref{Markov_transition})
 rather describes time reversible random walks over edges than over
nodes.
In other words, we are invited to consider random walks described
by the symmetric operator defined on the dual graph $G^\star$.

The Markov process (\ref{Markov_transition}) represents the
conservation of the traffic volume through the transport network,
while other solutions of (\ref{Lim_equilibrium}) are related to
the possible termination of travels along edges. If we denote the
number of neighbor edges the edge $e\in E$ has in the dual graph
$G^\star$ as $q_e=\deg_{G^\star}(e)$, then the simple substitution
shows that $w(e)=\sqrt{q_e}$ defines an eigenvector of the
symmetric Markov transition operator defined over the edges $E$
with eigenvalue 1. This eigenvector is positive and being properly
normalized determines the relative traffic volume through $e\in E$
at equilibrium.

Eq.(\ref{Markov_transition}) shows the essential role Markov's
 chains defined on edges play in equilibrium traffic modelling and emphasizes that
 the degrees of nodes are a key determinant of the transport networks properties.

The notion of traffic equilibrium had been introduced by J.G. Wardrop in
 \cite{Wardrop:1952} and then generalized in \cite{Beckmann:1956} to a
 fundamental concept of   network equilibrium. Wardrop's traffic
 equilibrium  is strongly tied to the human apprehension of space since
  it is required that all travellers have enough knowledge of the
  transport network they use. The human perception of places is not
  an entirely Euclidean one, but are rather related to the perceiving
  of the vista spaces (viewable spaces of streets and squares) as single units and to the
   understanding of the topological relationships between these vista spaces,
   \cite{Kuipers}.

The use of Eq.(\ref{Markov_transition}) also helps to clarify the inconsistency
of the  traditional axial technique widely implemented in space syntax theory.
Lines of sight are
 oversensitive to small
deformations of the grid, which leads to noticeably different axial graphs for systems
that should have similar configuration properties.
Long straight paths, represented by single lines, appear to be overvalued
compared to curved or sinuous paths as they are broken into a
number of axial lines that
creates an artificial differentiation between straight and curved or sinuous
 paths that have the same importance in the system \cite{Ratti}.
 Eq.(\ref{Markov_transition}) shows that the nodes of a
dual graph representing the  open spaces in the
spatial network of an urban environment should have an individual
meaning being an entity characterized by the certain traffic volume capacity.

Decomposition of city space into a complete set of
   intersecting open spaces characterized by the
traffic volume capacities produces a spatial network which we call
   the {\it dual} graph representation of a city.

\subsection{The processes associated with permutational automorphisms of the graph}
\label{subsec:automorphisms}
 \noindent

While analyzing a graph, whether it is primary or dual, we assign
the absolute scores to all nodes based on their properties with
respect to a transport process defined on that. Indeed, the nodes
of $G(V,E)$ can be weighted with respect to some  measure
 $m=\sum_{i\in V}\, m_i \,\delta_{ij},$ specified by a set of positive numbers $m_i> 0$.
The space $\ell^{\,2}(m)$ of square-summable functions with respect to the
   measure $m$ is a  Hilbert space $\mathcal{H}(V)$.

Among all measures which can be defined on $V$, the set of
normalized measures (or {\it densities}),
\begin{equation}
\label{denisty}
1\,=\,\sum_{i\in V}\, \pi_i\,\delta_{ij},
\end{equation}
 are of essential interest since they express the conservation of a
  quantity, and therefore may be relevant to a physical process.

The fundamental physical process defined on the graph is generated by the subset of its
automorphisms preserving the notion of connectivity of nodes. An automorphism is
 a mapping of the object to itself  preserving all of its structure. The set of
  all automorphisms of a graph forms a group, called the automorphism group.
For each graph $G(V,E)$, there  exists a unique,
up to permutations of rows and columns,
adjacency matrix $\mathbf{A}$,
 the $N\times N$ matrix defined by
 $A_{ij}=1$ if
  $i\sim j$,  and $A_{ij}=0$ otherwise.
As usual ${\bf A}$ is identified with a linear endomorphism of $C_0(G)$,
the vector space of all functions from $V$ into $\mathbb{R}$.
The degree of a node $i\in V$ is
therefore equal to
\begin{equation}
\label{condition}
k_i\,=\,\sum_{i\sim j}\,A_{ij}.
\end{equation}
Let us consider the set of all linear transformations
defined on the adjacency matrix,
\begin{equation}
\label{lin_fun}
Z\left({\bf A}\right)_{ij}\,=
\,\sum_{\,s,l=1}^N\, \mathcal{F}_{ijsl}\,A_{sl}, \quad \mathcal{F}_{ijsl}\,\in \,\mathbb{R},
\end{equation}
generated by the subset of automorphisms of the graph $G$.

The graph automorphisms are specified by the symmetric group $\mathbb{S}_N$
including all admissible permutations $p\in \mathbb{S}_N$
taking $i\in V$ to $p(i)\in V$. The representation of $\mathbb{S}_N$ consists of all
  $N\times N$ matrices ${\bf \Pi}_p,$ such that
  $\left({\bf \Pi}_p\right)_{i,\,p(i)}=1$, and $\left({\bf \Pi}_p\right)_{i,j}=0$
if $j\ne p(i).$

The function $Z\left({\bf A}\right)_{ij}$ should satisfy
\begin{equation}
\label{permut_invar}
{\bf \Pi}_p^\top\, Z\left({\bf A}\right)\,{\bf \Pi}_p\,=\,
Z\left({\bf \Pi}_p^\top\,{\bf A}\,{\bf \Pi}_p\right),
\end{equation}
for any $p\in \mathbb{S}_N$, and therefore entries of the tensor $\mathcal{F}$
must have the following symmetry property,
\begin{equation}
\label{symmetry}
\mathcal{F}_{p(i)\,p(j)\,p(s)\,p(l)}\,=\,
\mathcal{F}_{ijsl},
\end{equation}
for any $p\in \mathbb{S}_N$.  Since the action of the symmetric
group $\mathbb{S}_N$ preserves the conjugate classes of index
partition structures, any appropriate tensor $\mathcal{F}$
satisfying (\ref{symmetry}) can be expressed as a linear
combination of the following tensors: $ \left\{1,
\delta_{ij},\delta_{is},\delta_{il},\delta_{js},\delta_{jl},\delta_{sl},
\delta_{ij}\delta_{js},\delta_{js}\delta_{sl},\delta_{sl}\delta_{li},
\delta_{li}\delta_{ij},\right.\\
\left.
\delta_{ij}\delta_{sl},\delta_{is}\delta_{jl},\delta_{il}\delta_{js},
\delta_{ij}\delta_{il}\delta_{is} \right\}. $ Given a simple,
undirected graph $G$ such that $A_{ii}=0$ for any $i\in V$ then by
 substituting the above tensors into (\ref{lin_fun}) and taking
 account on symmetries we conclude that any arbitrary linear permutation invariant
 function must be of  the form
\begin{equation}
\label{lin_fun2}
Z\left({\bf A}\right)_{ij}\,=\,a_1+\delta_{ij}\,\left(a_2+a_3k_j\right)+
a_4\,A{}_{ij},
\end{equation}
with  $k_j=\deg_G(j)$ and $a_{1,2,3,4}$ arbitrary constants.

If we require that the
  linear function $Z$ preserves the notion of connectivity,
\begin{equation}
\label{conn_nodes}
k_i\,=\,\sum_{j\in V}\,Z\left({\bf A}\right)_{ij},
\end{equation}
it is clear that we should take $a_1=a_2=0$ (indeed, the
contributions $a_1N$ and $a_2$ are incompatible with
(\ref{conn_nodes})) and then obtain the relation for the remaining
constants, $1-a_3=a_4$. Introducing the new parameter $\beta\equiv
a_4>0$, we write (\ref{lin_fun2}) as follows,
\begin{equation}
\label{lin_fun3}
Z\left({\bf A}\right)_{ij}\,=\, (1-\beta)\,\delta_{ij} k_j +
\beta\,A_{ij}.
\end{equation}
If we express (\ref{conn_nodes})
in the form of the probability conservation relation, then the function
$Z\left({\bf A}\right)$ acquires  a probabilistic interpretation.
Substituting (\ref{lin_fun3}) back into (\ref{conn_nodes}), we obtain
\begin{equation}
\label{property}
\begin{array}{lcl}
1 & = & \sum_{j\in V}\, \frac{Z\left({\bf A}\right)_{ij}}{k_i} \\
  & = & \sum_{j\in V}\, (1-\beta)\,\delta_{ij} +\beta\,\frac{A_{ij}}{k_i}\\
 & = & \sum_{j\in V}\, T^{(\beta)}_{ij}.
\end{array}
\end{equation}
The operator $T^{(\beta)}_{ij}$ represents the generalized random
walk transition operator if $0<\beta\leq k^{-1}_{\max}$ where
$k_{\max}$ is the maximal node degree in the graph $G$. In the
random walks defined by $T^{(\beta)}_{ij}$, a random walker stays
in the initial vertex with probability $1-\beta$, while it moves
to another node randomly chosen among its nearest neighbors with
probability $\beta/k_i$. If we take $\beta=1$, then the operator
$T^{(\beta)}_{ij}$ describes the usual random walks discussed
extensively in the classical surveys \cite{Lovasz}-\cite{Aldous}.

Being defined on a connected aperiodic graph, the matrix $T^{(\beta)}_{ij}$
is a real positive stochastic matrix, and therefore, in accordance to the
 Perron-Frobenius theorem \cite{PerronFrobenius}, its maximal
 eigenvalue is 1, and it is simple. A left eigenvector
\begin{equation}
\label{pi}
\pi\,T^{(\beta)}\,=\,\pi
\end{equation}
 associated with the eigenvalue 1 has positive entries satisfying
  (\ref{denisty}). It is interpreted as a unique
 equilibrium state $\pi$ (stationary distribution of the random walk).
For
  any density $\sigma\in \mathcal{H}(V)$,
\begin{equation}
\label{limit}
\pi\,=\,\lim_{t\to\infty}\,\sigma\, T^{\,t}.
\end{equation}

\subsection{Transport network as Euclidean space}
\label{subsec:3}
 \noindent

Markov's operators on Hilbert space appear therefore as the natural
 language of complex network theory and space syntax theory in particular.
  Now we demonstrate that random walks embed connected undirected graphs
  into Euclidean space, in which distances and angles acquire the clear
  statistical interpretations.

The Markov operator $\widehat{T}$
is {\it self-adjoint} with respect to the normalized measure (\ref{denisty}) associated
to the stationary distribution of random walks $\pi$,
\begin{equation}
\label{s_a_analogue}
\widehat{T}=\,\frac 12
\left( \pi^{1/2}\,\, T\,\,
\pi^{-1/2}+\pi^{-1/2}\,\, T^\top\,\,
 \pi^{1/2}\right),
\end{equation}
where $T^\top$ is the transposed operator.
In the theory of
  random walks  defined on graphs \cite{Lovasz,Aldous} and in spectral
  graph theory \cite{Chung:1997}, basic properties of graphs are studied in
   connection with the  eigenvalues and eigenvectors of self-adjoint operators
   defined on them.
The orthonormal ordered set of real
 eigenvectors $\psi_i$, $i=1\ldots N$, of the symmetric operator $\widehat{T}$
 defines a basis in  $\mathcal{H}(V)$.

 In particular, the symmetric transition operator  $\widehat{T}$ of the
random walk
 defined on connected
   undirected graphs is
\begin{equation}
\label{transition_symm}
\widehat{T_{ij}}\,=\,
\left\{
\begin{array}{ll}
\frac 1{\sqrt{k_ik_j}},&  i\sim j, \\
0, & i\sim j.
\end{array}
\right.
\end{equation}
Its first eigenvector
    $\psi_1$ belonging to the largest eigenvalue $\mu_1=1$,
\begin{equation}
\label{psi_1}
\psi_1
\,\widehat{ T}\, =\,
\psi_1,
\quad \psi_{1,i}^2\,=\,\pi_i,
\end{equation}
describes the {\it local} property of nodes (connectivity),
since the stationary distribution of random walks  is
\begin{equation}
\label{stationary}
\pi_i\,=\,\frac{k_i}{2M}
\end{equation}
where $2M=\sum_{i\in V} k_i$.
The remaining eigenvectors,
 $\left\{\,\psi_s\,\right\}_{s=2}^N$, belonging to the eigenvalues
  $1>\mu_2\geq\ldots\mu_N\geq -1$ describe the {\it global} connectedness of the graph.
For example, the eigenvector corresponding to the second eigenvalue $\mu_2$
is used in spectral bisection of graphs. It is called the
Fiedler vector if related to the Laplacian matrix of a graph \cite{Chung:1997}.

Markov's symmetric transition operator $\widehat{T}$  defines a projection
 of any density $\sigma\in \mathcal{H}(V)$ on the eigenvector $\psi_1$ of the
  stationary distribution $\pi$,
\begin{equation}
\label{project}
\sigma\,\widehat{T}\,
=\,\psi_1 + \sigma_\bot\,\widehat{T},\quad \sigma_\bot\,=\,\sigma-\psi_1,
\end{equation}
in which $\sigma_{\bot}$ is the vector belonging to the orthogonal complement of
$\psi_1$.
In space syntax, we are interested in a comparison between the densities  with respect
 to random walks defined on the graph $G$.  Since all components $\psi_{1,i}>0$,
it is convenient to rescale the density $\sigma$ by dividing its
components by the components of $\psi_1$,
\begin{equation}
\label{rescaling}
\widetilde{\sigma_i}, =\,\frac{\sigma_i}{\psi_{1,i}}.
\end{equation}
Thus, it is clear that any two rescaled densities
$\widetilde{\sigma},\widetilde{\rho}\,\in\,\mathcal{H}$ differ
 with respect to random walks only by their dynamical components,
$$
\left(\widetilde{\sigma}-\widetilde{\rho}\right)\,
 \widehat{T}^t\,=\,\left(\widetilde{\sigma}_\bot -\widetilde{\rho}_\bot\right)\,
\widehat{T}^t,
$$
 for all $t>0$.
Therefore, we can define the
distance  $\|\ldots\|_T$ between any two densities established by random walks by
\begin{equation}
\label{distance}
\left\|\,\sigma-\rho\,\right\|^2_T\, =
\, \sum_{t\,\geq\, 0}\, \left\langle\, \widetilde{\sigma}_\bot -\widetilde{\rho}_\bot\,\left|\,\widehat{T}^t\,
\right|\, \widetilde{\sigma}_\bot -\widetilde{\rho}_\bot\,\right\rangle.
\end{equation}
 or, using the spectral
representation of $\widehat{T}$,
\begin{equation}
\label{spectral_dist}
\begin{array}{ll}
\left\|\sigma-\rho\right\|^2_T
&
=
\sum_{t\,\geq 0} \sum_{s=2}^N\, \mu^t_s \left\langle
\widetilde{\sigma}^\bot -\widetilde{\rho}^\bot|\psi_s\right\rangle\!\left\langle \psi_s
| \widetilde{\sigma}^\bot -\widetilde{\rho}^\bot\right\rangle
\\
 &
=\sum_{s=2}^N\,\frac{\left\langle\, \widetilde{\sigma}_\bot -\widetilde{\rho}_\bot\,|
\, \psi_s\,\right\rangle\!\left\langle\, \psi_s\,
| \,\widetilde{\sigma}_\bot -\widetilde{\rho}_\bot\,\right\rangle}{\,1\,-\,\mu_s\,},
\end{array}
\end{equation}
where we have used  Dirac's bra-ket notations especially
convenient for working with inner products and
rank-one
operators in Hilbert space.

If we introduce a new inner product for
densities $\sigma,\rho \in\mathcal{H}(V)$
by
\begin{equation}
\label{inner-product}
\left(\,\sigma,\rho\,\right)_{T}
\,= \,  \sum_{s=2}^N
\,\frac{\,\left\langle\,  \widetilde{\sigma}_\bot\,|\,\psi_s\,\right\rangle\!
\left\langle\,\psi_s\,|\, \widetilde{\rho}_\bot \right\rangle}{\,1\,-\,\mu_s\,},
\end{equation}
then (\ref{spectral_dist}) is nothing else but
\begin{equation}
\label{spectr-dist2}
\left\|\,\sigma-\rho\,\right\|^2_T\, =
\left\|\,\sigma\,\right\|^2_T +
\left\|\,\rho\,\right\|^2_T  -
2 \left(\,\sigma,\rho\,\right)_T,
\end{equation}
 where
\begin{equation}
\label{sqaured_norm}
\left\|\, \sigma\,\right\|^2_T\,=\,
\,\sum_{s=2}^N \,\frac{\left\langle\,  \widetilde{\sigma}_\bot\,|\,\psi_s\,\right\rangle\!
\left\langle\,\psi_s\,|\, \widetilde{\sigma}_\bot\, \right\rangle}{\,1\,-\,\mu_s\,}
\end{equation}
is the square
of the norm of  $\sigma\,\in\, \mathcal{H}(V)$ with respect to
random walks defined on the graph $G$.

We finish the description of the $(N-1)$-dimensional Euclidean
space structure of $G$ induced by
  random walks by mentioning that
given two densities $\sigma,\rho\,\in\, \mathcal{H}(V),$ the
angle between them can be introduced in the standard way,
\begin{equation}
\label{angle}
\cos \,\angle \left(\rho,\sigma\right)=
\frac{\,\left(\,\sigma,\rho\,\right)_T\,}
{\left\|\,\sigma\,\right\|_T\,\left\|\,\rho\,\right\|_T}.
\end{equation}
Random walks embed connected undirected graphs into the Euclidean
space $\mathbb{R}^{N-1}$. This embedding  can be used in order to compare
 nodes
and to construct
 the optimal coarse-graining
representations.

Namely, in accordance to (\ref{sqaured_norm}), the density $\delta_i$, which
equals 1 at $i\in V$
and zero otherwise,
acquires the norm $\left\|\,\delta_i\,\right\|_T$
associated to random walks defined on $G$.
In the theory of random walks \cite{Lovasz},
 its square,
\begin{equation}
\label{norm_node}
\left\|\,\delta_i\,\right\|_T^2\, =\,\frac 1{\pi_i}\,\sum_{s=2}^N\,
\frac{\,\psi^2_{s,i}\,}{\,1-\mu_s\,},
\end{equation}
 gets
a clear probabilistic interpretation
expressing
the {\it access time} to a target node
quantifying the expected number
of  steps
required for a random walker
to reach the node
$i\in V$ starting from an
arbitrary
node  chosen randomly
among all other
nodes  with respect to
the stationary distribution $\pi$.

The Euclidean distance between any two nodes of the graph $G$
calculated as the distance (\ref{spectral_dist}) between the densities $\delta_i$ and
$\delta_j$
induced by random walks,
\begin{equation}
\label{commute}
K_{i,j}\,=\,\left\|\, \delta_i-\delta_j\,\right\|^2_T\,=\, \sum_{s=2}^N\,\frac 1{1-\mu_s}\left(\frac{\psi_{s,i}}{\sqrt{\pi_i}}-\frac{\psi_{s,j}}{\sqrt{\pi_j}}\right)^2,
\end{equation}
quantifies the {\it commute time} in theory of random walks being
equal to the expected number of steps required for a random
walker starting at $i\,\in\, V$ to visit $j\,\in\, V$ and then to
return back to $i$,  \cite{Lovasz}.

It is important to mention that
the cosine of an angle calculated in accordance to
 (\ref{angle}) has the structure of
Pearson's coefficient of linear correlations
 that reveals it's natural
statistical interpretation.
Correlation properties of flows
of random walkers
passing by different paths
 have been remained beyond the scope of
previous  studies devoted to complex
networks and random walks on graphs.
The notion of angle between any two nodes of the
graph arises naturally as soon as we
become interested in
the strength and direction of
a linear relationship between
two random variables,
the flows of random walks moving through them.
If the cosine of an angle (\ref{angle}) is 1
(zero angles),
there is an increasing linear relationship
between the flows of random walks through both nodes.
Otherwise, if it is close to -1 ($\pi$ angle),
  there is
a decreasing linear relationship.
The  correlation is 0 ($\pi/2$ angle)
if the variables are linearly independent.
It is important to mention that
 as usual the correlation between nodes
does not necessary imply a direct causal
relationship (an immediate connection)
between them.

\section{Examples: Petersen graph and the network of Venetian Canals}
\label{sec:Examples}
\noindent

\begin{figure}[ht]
 \noindent
\begin{center}
\epsfig{file=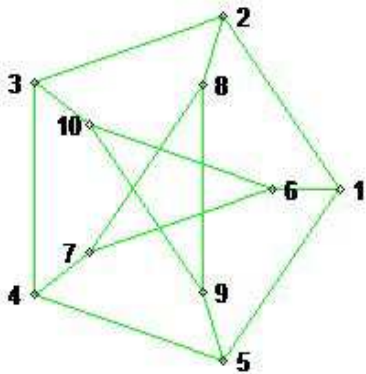,  angle= 0,width =5cm, height =5cm}
  \end{center}
\caption{\small The Petersen graph.}
\label{Fig1}
\end{figure}

\begin{figure}[ht]
 \noindent
\begin{center}
\epsfig{file=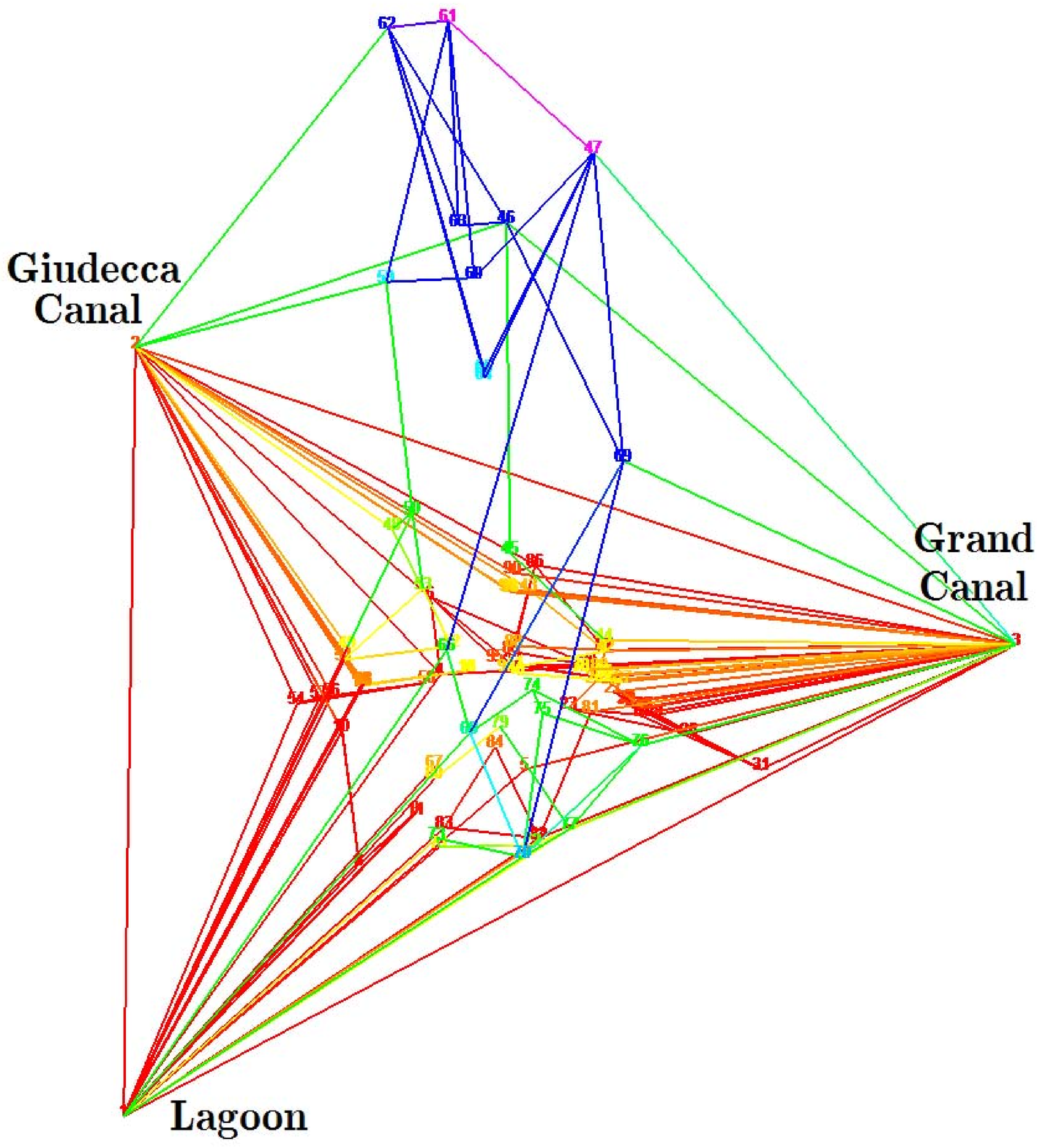,  angle= 0,width =8cm, height =8cm}
  \end{center}
\caption{\small The dual graph representation of the spatial network of 96 Venetian canals }
\label{Fig2}
\end{figure}

In the present section, we construct  the Euclidean embedding of
two graphs. One graph we study is the Petersen graph of  10 nodes
(see Fig.~\ref{Fig1}). Another example is the spatial network of
96 Venetian canals which serve the function of roads in the
ancient city that stretches across 122 small islands (see
Fig.~\ref{Fig2}). While identifying a canal over the plurality of
water routes on the city map of Venice, the canal-named approach
has been used, in which two different arcs of the city canal
network were assigned to the same identification number provided
they have the same name.
The Petersen graph is a regular graph, $k_i=3$, $i=1,\ldots 10$,
so that $\sum_{i}k_i=30$, and the stationary distribution of
random walks is uniform, $\pi^{(\mathrm{Pet})}_i=0.1$. The
spectrum of the random walk transition operator
(\ref{transition_symm}) contains the Perron eigenvalue $1$ which
is simple, then the eigenvalue $1/3$ of multiplicity $5$, and
$-2/3$ of multiplicity $4$. Therefore, there are just $3$ linearly
independent eigenvectors, and two eigensubspaces for which the
orthonormal basis vectors can be calculated, so that the resulting
matrix of eigenvectors and basis vectors which we use in
(\ref{norm_node}-\ref{commute}) always has full column dimension.
Random walks embed the Petersen graph into a $9$-dimensional
Euclidean space, in which all nodes have equal norm
(\ref{norm_node}), $\left\| i\right\|_T= 3.14642654$ that means
the expected number of  steps a random walker starting from a node
chosen randomly with probability $p=0.1$ reaches any node in the
Petersen graph equals $9.9$.
\begin{figure}[ht]
 \noindent
\begin{center}
\epsfig{file=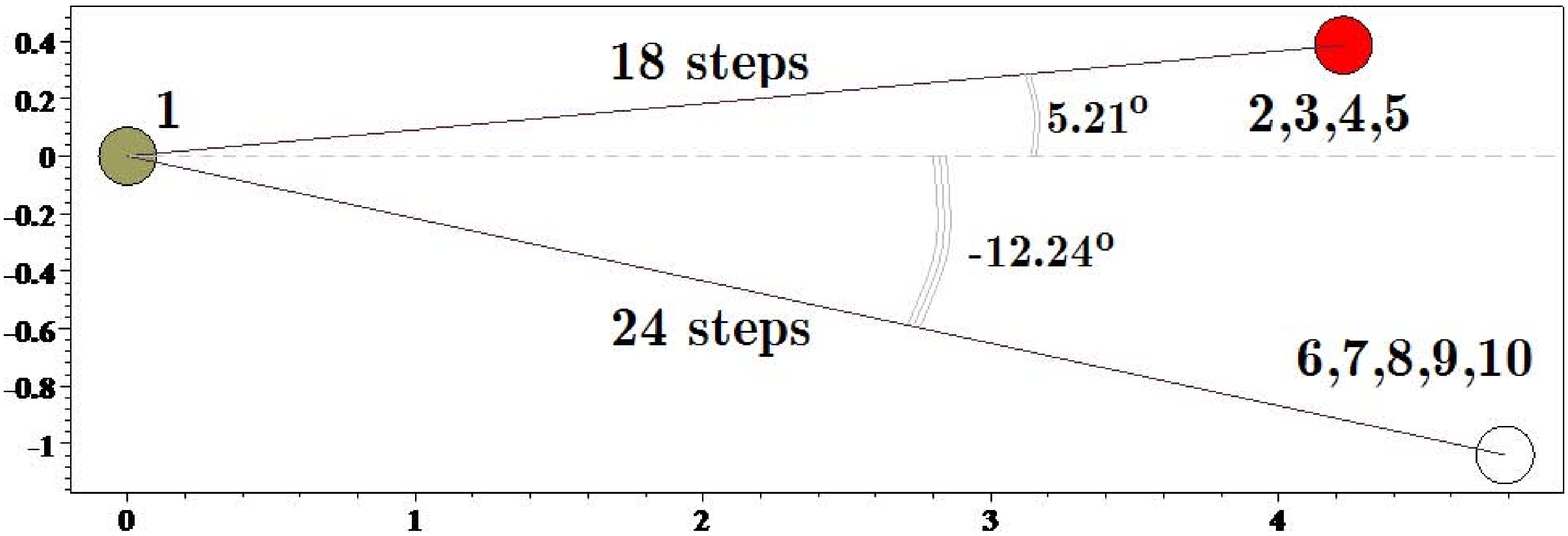,  angle= 0,width =10cm, height =3.5cm}
  \end{center}
\caption{\small The Euclidean space embedding of the Petersen
graph drawn with respect to the node $\# 1$.}
\label{Fig3}
\end{figure}
Indeed, the structure of $9$-dimensional vector space  induced by
random walks defined on the Petersen graph cannot be represented
visually, however if we choose a node as a point of reference, we
can draw its 2-dimensional projection  by arranging other nodes at
the distances calculated accordingly to (\ref{commute}) and under
the angles found from (\ref{angle}) they are with respect to the
chosen reference node (see Fig.~\ref{Fig3}).

It is expected that a random walker starting at $\# 1$
visits any periphery node ($\# 2,3,4,5$)
and then returns back in 18 random steps, while it is required 24
random steps in order to visit any internal node in the deep of the graph
($\# 6,7,8,9,10$). It is also obvious that while the
linear relationship
between the  random walks flows
through  $\# 1$ and those through the periphery nodes
is positive, it is negative with respect to
 the flows passing through the internal nodes.
Due to the symmetry of the Petersen graph, the figure displayed on Fig.~\ref{Fig3}
is essentially the same if we draw it with respect to any other periphery node
 ($\# 2,3,4,5$). It is also important to note that it turns to be mirror-reflected
if we draw it with respect to any internal node
($\# 6,7,8,9,10$). Therefore, we can conclude that the
 Petersen graph contains two groups of nodes, at the
 periphery and in deep, which appears to be "a quarter" higher segregated
  from each other (18 random steps vs. 24 random steps).
It is clear that the $9-$dimensional embedding of the Petersen graph
into Euclidean space is characterized by the highest degree of symmetry.

\begin{figure}[ht]
 \noindent
\begin{center}
\epsfig{file=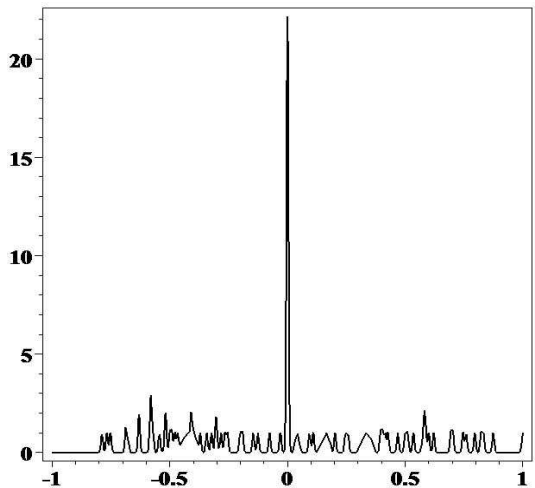,  angle= 0,width =7cm, height =6.5cm}
  \end{center}
\caption{\small The spectrum of the Markov transition operator
(\ref{transition_symm}) defined on the spatial network of Venetian canals.}
\label{Fig4}
\end{figure}

The graph representation of the spatial network of Venetian canals
(Fig~\ref{Fig2}) is much more complicated than the Petersen graph.
The graph is far from being regular, so that the stationary
distribution of random walks defined on it is not uniform.
In \cite{cities:2007}, we have discussed that it is not evident
that the degree distributions in compact urban patterns and in
Venice, in particular, follow a power law.
The spectrum of the Markov transition operator
(\ref{transition_symm}) defined on that is presented in Fig.~\ref{Fig4}.
The matrix (\ref{transition_symm}) for the canal network in Venice
is strongly defective, In particular, it contains the eigenvalue
$\mu=0$ of multiplicity 22. This degeneracy indicates the presence
of the complete bipartite subgraph in the spatial network of
Venice shown in Fig.~\ref{Fig2}.
\begin{figure}[ht]
 \noindent
\begin{center}
\epsfig{file=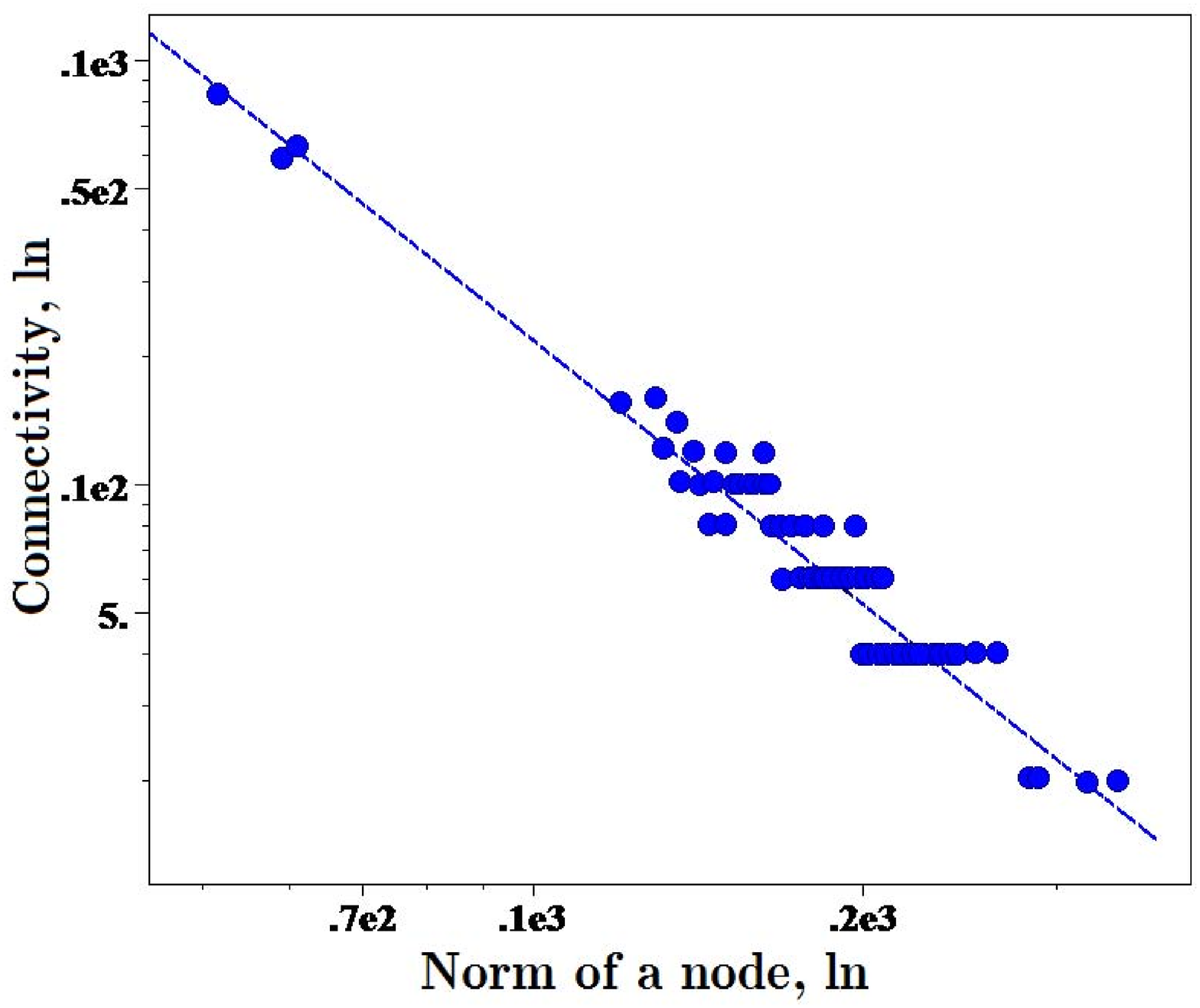,  angle= 0,width =7cm, height =6.5cm}
  \end{center}
\caption{\small The scatter plot of the connectivity vs. the norm a node in
the dual graph representation of 96 Venetian canals acquires with respect
to random walks.
Three data points characterized by the shortest access times represent the
main water routes of Venice: the Lagoon of Venice, the Giudecca canal, and
the Grand canal. Four data points of the worst accessibility are for the
canal subnetwork of Venetian Ghetto.
The slope of the regression
 line equals 2.07.}
\label{Fig5}
\end{figure}
The norms of canals with respect to random walks are different and
 scales with their connectivity (see Fig.~\ref{Fig5}). The notion of spatial
segregation acquires a statistical interpretation
with respect to random walks by means of (\ref{norm_node}).
In urban
spatial networks encoded by their dual graphs, the access times
(\ref{norm_node}) strongly
vary  from one open space
to another and could be very large for
statistically segregated spaces. Three data points characterized by
the shortest access times shown in Fig.~\ref{Fig5} are due to the Lagoon
of Venice, the Giudecca canal, and
the Grand canal - the most central water routes in the city canal network.
Four data points characterized by the worst accessibility represent the
canal subnetwork of Venetian Ghetto.
The slope of the regression
 line equals 2.07.

\begin{figure}[ht]
 \noindent
\begin{center}
\epsfig{file=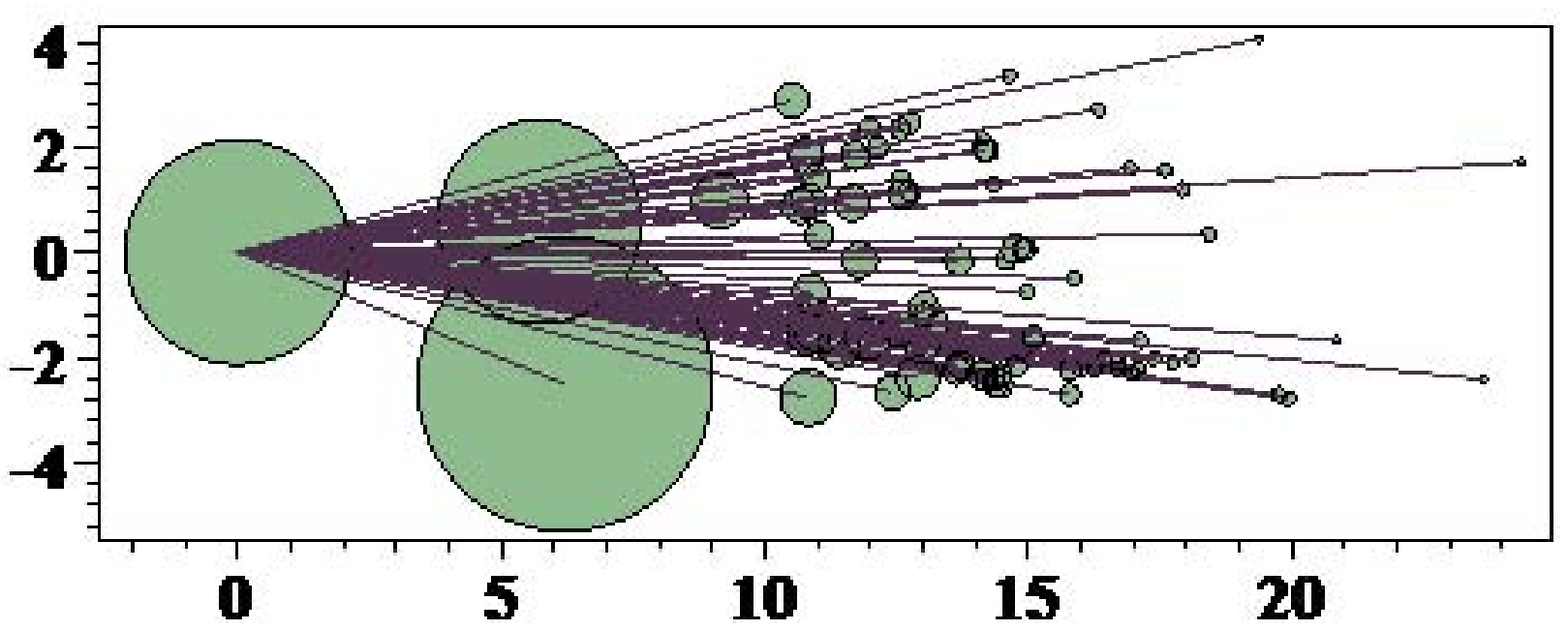,  angle= 0,width =10cm, height =5cm}
  \end{center}
\caption{\small The 2-dimensional projection of the 95-dimensional
 Euclidean spaces associated to random walks defined on the city
 canal network built from the perspective of the Grand canal of Venice
chosen as the origin. The labels of the
horizontal axes display the expected number of random walk steps.
The labels of the
vertical axes show the degree of nodes (radiuses of the disks). }
\label{Fig6}
\end{figure}
The 2-dimensional projection
of the  Euclidean space of 96 Venetian canals  set up by random walks
drawn for the
 the Grand Canal of Venice (the point $(0,0)$) is shown in Fig.~\ref{Fig6}.
Nodes of the dual graph representation of
 the canal network in Venice
are shown by disks with radiuses taken equal to the degrees of the
nodes. All distances between the chosen origin and other nodes of the
graph (Fig.~\ref{Fig2}) have been calculated in accordance to
(\ref{commute}) and (\ref{angle}) has been used in
order to compute angles between nodes.
Canals negatively correlated with the Grand Canal of Venice are
set under negative angles (below the horizontal), and under
positive angles (above the horizontal) if otherwise.

It is evident from Fig.~\ref{Fig6}
that disks of smaller radiuses
demonstrate a clear
tendency to be located far
away from the origin
being characterized by
the excessively long commute times with the reference point (the Grand canal of Venice),
while the large disks which stand in Fig.~\ref{Fig6} for the main water routes
are settled in the closest proximity to the
origin that intends an immediate access to them.

\section{Discussion and Conclusion}
\label{sec:Discussion}
\noindent

In the present paper, we have developed a self-consistent approach
to modelling of complex networks. We have discussed the possible
creation algorithm (see Sec.~\ref{sec:Cameo}) which refers to the idea of
distribution of resources, including land, water, minerals, fuel
and wealth in general rather than to the popularity driven
preferential attachment approach proposed in \cite{BA,BA2}. We
have also demonstrated that random walks are the effective tool
for investigation of the graph structure since they embed a
connected graph into Euclidean space, in which the distances and
angles acquire the statistical interpretations.

Probably, the most important conclusion of space syntax theory is that
the adequate
 level of the positive relationship between
the connectivity
of city spaces and their integration property (vs. segregation)
called intelligibility
encourages
  peoples way-finding abilities \cite{Hillier:1999}.
Intelligibility of Venetian
  canal network reveals itself quantitatively in the scaling of the
   norms of nodes with connectivity shown in Fig.~\ref{Fig5} and
   qualitatively in the tendency of smaller disks to be located on the
    outskirts of the Venetian space syntax displayed in Fig~\ref{Fig6}.

\section*{Acknowledgment}
\label{Acknowledgment}
\noindent

The support from the Volkswagen Foundation (Germany) in the
framework of the project "{\it Network formation rules, random set
graphs and generalized epidemic processes}" is gratefully
acknowledged.

\end{document}